%% file: main.tex
\documentclass[sigconf,natbib=true]{acmart}
\AtBeginDocument{%
  }
\usepackage{etoolbox}

\makeatletter
\newcommand{\customsize}{\@setfontsize\customsize{9.5pt}{11pt}} 
\makeatother

\usepackage{graphicx}
\usepackage{float}
\usepackage{subfigure}
\usepackage{wrapfig}
\usepackage{float}
\usepackage[utf8]{inputenc} 
\usepackage[T1]{fontenc}    
\usepackage{url}            
\usepackage{booktabs}       
\usepackage{amsfonts}       
\usepackage{nicefrac}       
\usepackage{microtype}      
\usepackage{xcolor}         
\usepackage{ragged2e} 
\usepackage{booktabs,makecell, multirow, tabularx}
\usepackage{colortbl}
\usepackage{algorithmic}
\usepackage[ruled,linesnumbered,vlined]{algorithm2e}

\usepackage{enumitem}
\usepackage{verbatim}
\usepackage{CJKutf8}

\newcommand{\ie}{\textit{i.e., }}
\newcommand{\eg}{\textit{e.g., }}

\definecolor{mygray}{rgb}{0.9, 0.9, 0.9}
\usepackage{tcolorbox} 
\usepackage{graphicx} 

\copyrightyear{2025}
\acmYear{2025}
\setcopyright{acmlicensed}\acmConference[SIGIR '25]{Proceedings of the 48th
International ACM SIGIR Conference on Research and Development in
Information Retrieval}{July 13--18, 2025}{Padua, Italy}
\acmBooktitle{Proceedings of the 48th International ACM SIGIR Conference on
Research and Development in Information Retrieval (SIGIR '25), July 13--18,
2025, Padua, Italy}
\acmDOI{10.1145/3726302.3729890}
\acmISBN{979-8-4007-1592-1/2025/07}
\begin{document}

\title{Addressing Missing Data Issue for Diffusion-based Recommendation}

\author{Wenyu Mao}
\affiliation{%
  \institution{University of Science and Technology of China}
  \city{Anhui, Hefei}
  \country{China}}
\email{wenyumao2@gmail.com}

\author{Zhengyi Yang}
\affiliation{%
  \institution{University of Science and Technology of China}
 \city{Anhui, Hefei}
  \country{China}}
\email{yangzhy@mail.ustc.edu.cn}

\author{Jiancan Wu*}\thanks{* corresponding authors}
\affiliation{%
  \institution{University of Science and Technology of China}
 \city{Anhui, Hefei}
  \country{China}}
  \email{wujcan@gmail.com}

\author{Haozhe Liu}
\affiliation{%
  \institution{University of Science and Technology of China}
 \city{Anhui, Hefei}
  \country{China}}
  \email{liuhz0803@gmail.com}
\author{Yancheng Yuan}
\affiliation{%
  \institution{The Hong Kong Polytechnic University}
 \city{Hong Kong}
  \country{China\\\customsize {yancheng.yuan@polyu.edu.hk}}}

\author{Xiang Wang}
\affiliation{%
  \institution{University of Science and Technology of China}
 \city{Anhui, Hefei}
  \country{China}\\\customsize{xiangwang1223@gmail.com}}

\author{Xiangnan He*}
\affiliation{%
  \institution{University of Science and Technology of China}
 \city{Anhui, Hefei}
  \country{China}\\\customsize{xiangnanhe@gmail.com}}

\renewcommand{\shortauthors}{Wenyu Mao et al.}

\begin{abstract}
Diffusion models have shown significant potential in generating oracle items that best match user preference with guidance from user historical interaction sequences. However, the quality of guidance is often compromised by unpredictable missing data in observed sequence, leading to suboptimal item generation. Since missing data is uncertain in both occurrence and content, recovering it is impractical and may introduce additional errors. To tackle this challenge, we propose a novel dual-side \textbf{T}hompson sampling-based \textbf{D}iffusion \textbf{M}odel (TDM), which simulates extra missing data in the guidance signals and allows diffusion models to handle existing missing data through extrapolation. To preserve user preference evolution in sequences despite extra missing data, we introduce Dual-side Thompson Sampling to implement simulation with two probability models, sampling by exploiting user preference from both item continuity and sequence stability. TDM strategically removes items from sequences based on dual-side Thompson sampling and treats these edited sequences as guidance for diffusion models, enhancing models' robustness to missing data through consistency regularization. Additionally, to enhance the generation efficiency, TDM is implemented under the denoising diffusion implicit models to accelerate the reverse process. Extensive experiments and theoretical analysis validate the effectiveness of TDM in addressing missing data in sequential recommendations. Our data
and code is available at \url{https://github.com/maowenyu-11/TDM}.

\end{abstract}



\maketitle
\input{1_introduction.tex}
\input{5_related_work}

\input{2_problem_formulation.tex}
\input{3_methodology.tex}

\input{4_experiments.tex}
\input{6_conclusion.tex}
\balance
\bibliographystyle{ACM-Reference-Format}
\bibliography{reference}

\end{document}

%% file: 1_introduction.tex
\section{Introduction}
Sequential Recommendation \citep{SASRec,mao2025reinforced, LLaRa, ilora} is to predict the next item that aligns with a user's preferences based on his/her historical interaction sequence. Unlike conventional studies \citep{CL4Rec, CLRDS, session} that learn to classify target items from sampled negatives, recent studies \citep{SVAE, yiyan_genout, gen-retrival} shift towards generating oracle items \citep{DreamRec} with generative models that best match user preference.
A promising direction is employing diffusion models \citep{DreamRec, DiffuRec, Diffris}, which add noise to the next items and iteratively denoise them toward oracle items, guided by interaction history conditions.

\begin{figure}
  \centering
  \includegraphics[width=0.48\textwidth]{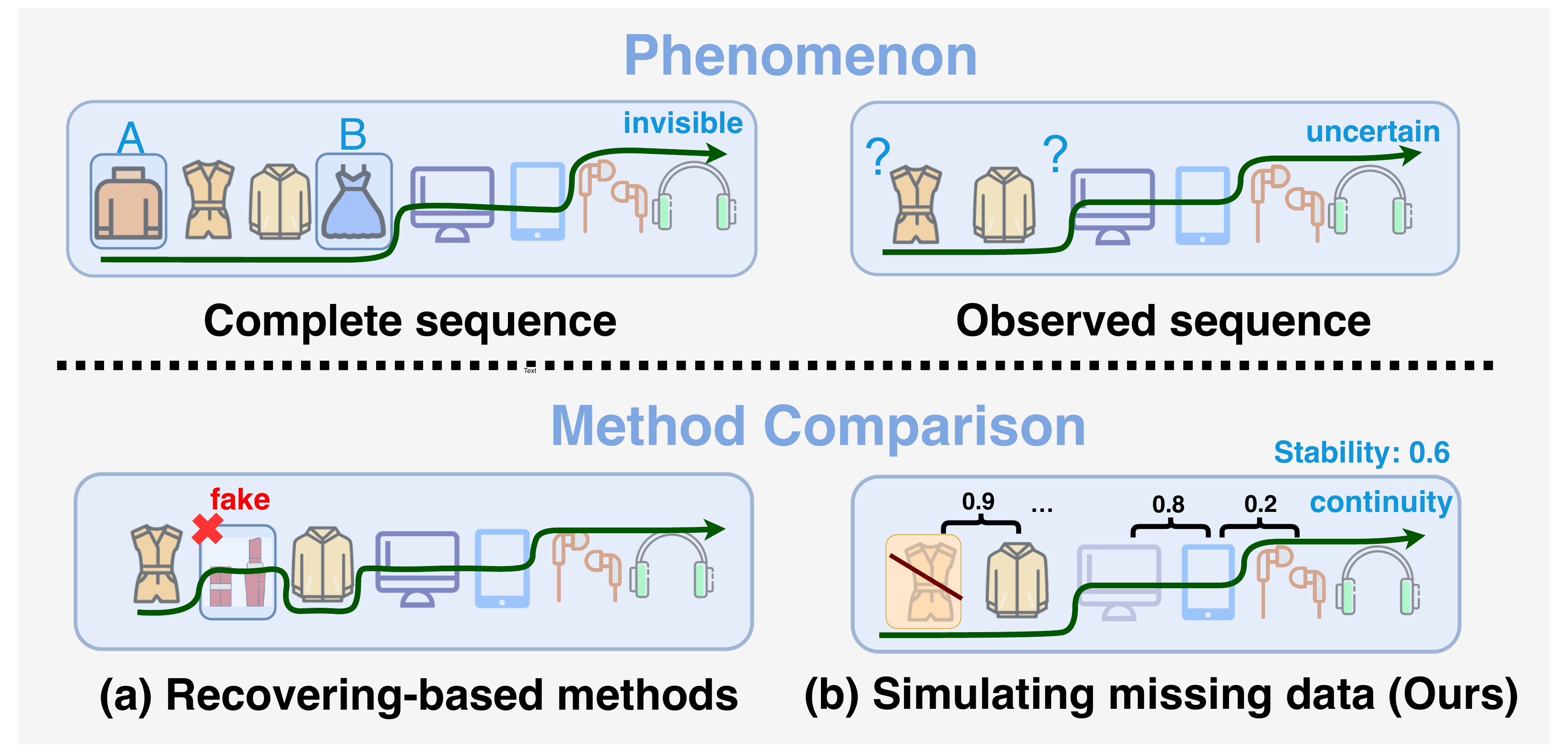}
   \vspace{-12pt}
  \caption{Phenomenon of uncertain missing data in sequences and the method comparison to address it. The green curve represents the evolution of user preference over time.}
 \vspace{-14pt}
  \label{fig: methods_comparison}
\end{figure}

However, we argue that diffusion models' ability to generate oracle items is largely constrained by missing data in the interaction history. Typically, user interaction histories are only partially observed, with missing data occurring unpredictably \citep{missing}. Consider the case as illustrated in Figure \ref{fig: methods_comparison}, the recommender system might only observe a partial sequence, with items $A$ and $B$ missing due to various factors, such as privacy concerns \citep{adapting} or technical limitations \citep{adversarial}. Consequently, diffusion models may be misled by the unreliable guidance signal from the observed sequence and generate suboptimal oracle items. 
Usually, missing data is \textbf{uncertain} in the observed sequence \citep{uncertainty1, uncertainty2},
as it is hard to infer where the missing occurs and what content it might be due to the \textbf{invisibility} of complete sequences. Thus, leading approaches that aim to recover missing data \cite{imputation1,imputation2} and complete observed sequences \citep{STEAM, SSDRec}, may introduce additional errors or distort user preference accidentally, as shown in Figure \ref{fig: methods_comparison}.

To address this challenge, we propose a dual-side \textbf{T}hompson sampling-based \textbf{D}iffusion \textbf{M}odel (TDM), which simulates extra missing data in the guidance signals rather than recovering existing one. Such simulation enables diffusion models to address missing data in real-world scenarios through extrapolation, suggesting that if diffusion models can handle simulated missing data, they can also manage real missing data.
To simulate while maintaining the underlying preference evolution (the changing of user preference over time) in sequences, we introduce the Dual-side Thompson Sampling \citep{Thompson} (DTS) strategy, which samples items to remove by exploiting known user preference. Specifically, DTS hires two probability models --- one operating at the local item side and the other at the global sequence side --- to capture dynamic preference:

\begin{itemize}[leftmargin=*]
\item  The local model depicts the continuity between adjacent items, reflecting shifts in user preference. As depicted in Figure \ref{fig: methods_comparison}, a high continuity score of 0.9 indicates a coherent preference for clothes, while a low score of 0.2 suggests a preference shift from mobile phones to headphones. 
\item The global model evaluates the stability of each entire sequence, by calculating the entropy of continuity score distribution --- a high stability score indicates a stable preference, whereas a low score reflects a volatile preference. For instance, the sequence in Figure \ref{fig: methods_comparison} experienced two significant fluctuations, resulting in a stability score of 0.6.
\end{itemize}
High-continuity items in high-stability sequences are more likely to be removed by DTS, which has little impact on the underlying preference evolution in sequences. This is evidenced by the consistent green curves between the observed sequence and the one edited by our method, as illustrated in Figure \ref{fig: methods_comparison}.
Then, we treat such edited sequences as guidance for diffusion models to generate the oracle items, which can achieve consistency regularization \citep{consistency} and endows diffusion models with insensitivity to preference-preserving perturbations (\ie simulated missing data).

To further improve the efficiency, we utilize denoising diffusion implicit models \citep{DDIM} rather than denoising diffusion probabilistic models \citep{DDPM} to generate oracles, which can accelerate the generation during inference.
To validate, we provide a theoretical analysis of extrapolation and consistency regularization for TDM. Additionally, extensive experiments demonstrate that TDM outperforms multiple leading models in sequential recommendations.
Our key contributions are as follows:
\begin{itemize}[leftmargin=*]
\item We propose TDM to simulate extra missing data in the guidance signals, enabling diffusion models to handle existing missing data through extrapolation and consistency regularization.

\item We introduce Dual-side Thompson Sampling to implement simulation based on user preference and apply denoising diffusion implicit models to accelerate generation.

\item Theoretical analysis and extensive experiments validate the effectiveness of TDM in addressing missing data.
\end{itemize}

%% file: 5_related_work.tex
\section{Related Work}
In this section, we provide a review of missing data in recommendation and generative recommenders.

\vspace{5pt}
\noindent\textbf{Missing Data in Sequential Recommendation}
\label{missing_data}
\citep{missing, counterCLR} refers to the absence of user behaviors in interaction sequences due to complex factors, which may lead to sequences partially observed and unreliable. To address this problem, recovering-based methods \citep{PDRec, SSDRec, STEAM} have become mainstream, aiming to recover the complete sequence through imputation \citep{imputation1, imputation2, imputation3}. For instance, PDRec \citep{PDRec} leverages diffusion models to generate supplement items to the observed sequences. SSDRec \citep{SSDRec} augments the interaction sequence by “insert” operations. Here we emphasize the uncertain nature \citep{uncertainty1, uncertainty2} of missing data, which poses challenges for recovering-based methods due to the invisibility of complete sequences as labels. Thus we propose simulating missing data instead of recovering it, enabling diffusion models' extrapolation to address real missing data. 

\vspace{5pt}
\noindent\textbf{Diffusion-based generative recommender} aims to generate oracle items that best match user preference,
offering distinct advantages over discriminative recommenders that learn to classify target items from sampled negatives, particularly in sequential recommendation tasks. 
Since GANs and VAEs are limited in the stability and quality of generation, diffusion models have emerged as a promising technique, excelling at modeling complex data distributions and generating oracle items \citep{DreamRec, Diffris, graphdiff, DiQDiff}. For example, DiffuRec \citep{DiffuRec}, DreamRec \citep{DreamRec}, and DimeRec \cite{DimeRec} generate the next items directly by corrupting them with noise and denoising based on the historical sequence. Additionally, DiffuASR \citep{DiffuASR},  DiffKG \cite{DiffKG}, and CaDiRec \citep{cadirec} enhance the traditional recommenders by generating sequences or items with diffusion models as data augmentation. Moreover, RecDiff \cite{RecDiff} and DDRM \cite{DDRM} leverage the denoising ability of Diffusion models to improve recommenders' robustness against noisy feedback. In our work, we emphasize the quality of diffusion models' guidance, tackling challenges posed by uncertain missing data and enhancing the robustness through consistency regularization.

%% file: 2_problem_formulation.tex
\section{Preliminaries}

In this section, we first detail denoising diffusion implicit models \citep{DDIM}, which can accelerate the reverse process of generation.
We then introduce Thompson sampling.
Finally, we formulate the task of generative sequential recommendation.

\subsection{Denoising Diffusion Implicit Models}
\label{ddim}
Denoising Diffusion Implicit Models \citep{DDIM} is designed to generate samples faster than the original denoising diffusion probabilistic models. Here we explain its forward and reverse processes.

\vspace{5pt}
\noindent\textbf{Forward process:}
Unlike denoising diffusion probabilistic models \citep{DDPM}, the forward process of denoising diffusion implicit models is not restricted as a Markovian chain, which enables it to denoise with fewer steps.
Given an input data sample $\mathbf{x}^0\sim q(\mathbf{x}^0)$, the forward diffusion process can be defined as: 
$
q(\mathbf{x}^t|\mathbf{x}^0)=\mathcal{N}(\mathbf{x}^t;\sqrt{\alpha_t}\mathbf{x}^0,(1-\alpha_t)\mathbf{I})
$,
where $t \in[1, \ldots, T]$
represents the diffusion step, $\left[\alpha_1, \ldots, \alpha_T\right]$ denotes a variance schedule. We can have: $\mathbf{x}^t=\sqrt{\alpha_t}\mathbf{x}^0+\sqrt{1-\alpha_t}\boldsymbol{\epsilon}$.

\vspace{5pt}
\noindent\textbf{Reverse process:}
Given $\mathbf{x}^T\sim\mathcal{N}(\mathbf{0},\mathbf{I})$, denoising diffusion implicit models eliminate the noises to recover $\mathbf{x}^0$ step by step.
Formally, the reverse process from $\mathbf{x}^t$ to $\mathbf{x}^{t-1}$ is: 
\begin{gather}
p_{\theta,\sigma}(\mathbf{x}^{t-1}|\mathbf{x}^t)=\mathcal{N}\left(\sqrt{\alpha_{t-1}}\left.\hat{\mathbf{x}}^0+\sqrt{1-\alpha_{t-1}-\sigma_t^2}\frac{\mathbf{x}^t-\sqrt{\alpha_t}\left.\hat{\mathbf{x}}^0\right.}{\sqrt{1-\alpha_t}},\sigma_t^2I\right)\right.,\\
\hat{\mathbf{x}}^0=(\mathbf{x}^t-\sqrt{1-\alpha_t}\boldsymbol{\epsilon}_\theta(\mathbf{x}^t))/\sqrt{\alpha_t} .
\label{x_0}
\end{gather}
When $\sigma_t=0$, it becomes a deterministic process.
$\boldsymbol{\epsilon}_{\theta}$ denotes the denoising model (\eg U-Net \citep{Unet} or Transformer \citep{Transformer}) parameterized by $\theta$, which is trained to approximate the data distribution $q\left(\mathbf{x}^0 \right)$ by maximizing the evidence lower bound of the $\log$-likelihood $\log p_\theta\left(\mathbf{x}^0\right)$. The training loss can be derivated as \citep{DDIM}:
\begin{gather}
\label{loss_pre}
\mathcal{L}=\sum_{t=1}^T\frac1{2d\sigma_t^2\alpha_t}\mathbb{E}_{\mathbf{x}^0,\boldsymbol{\epsilon}}\left[\left\|\boldsymbol{\epsilon}_\theta(\sqrt{\alpha_t}\mathbf{x}^0+\sqrt{1-\alpha_t}\boldsymbol{\epsilon},t)-\boldsymbol{\epsilon}\right\|_2^2\right],
\end{gather}
where $\boldsymbol{\epsilon}_\theta(\sqrt{\alpha_t}\mathbf{x}^0+\sqrt{1-\alpha_t}\boldsymbol{\epsilon},t)$ is the output of the denoising network to predict
the noises $\boldsymbol{\epsilon}$ that add in the forward process, $d$ is the dimension of $\mathbf{x}^0$.

\vspace{5pt}
\noindent\textbf{Acceleration:} 
\label{accelerated}
Since the denoising objective $\mathcal{L}$ is independent of a specific forward process as long as $q(\mathbf{x}^t|\mathbf{x}^0)$ is fixed \citep{DDIM}, we can redefine the {non-Markovian} forward process with a subsequence $[\tau_1,\tau_2,\cdots,\tau_S]$ from $[1, \ldots, T]$ as: $
q(\mathbf{x}^{\tau_s}|\mathbf{x}^0)=\mathcal{N}(\mathbf{x}^{\tau_s};\sqrt{\alpha_{\tau_s}}\mathbf{x}^0,(1-\alpha_{\tau_s})\mathbf{I})$.
Then, the reverse process can be reformulated as:
\begin{align}
\label{reverse}
\mathbf{x}^{\tau_{s-1}}&=\sqrt{\alpha_{\tau_{s-1}}}\Big(\frac{\mathbf{x}^{\tau_{s}}-\sqrt{1-\alpha_{\tau_{s}}}\boldsymbol{\epsilon}_{\theta}(\mathbf{x}^{\tau_{s}},\tau_{s})}{\sqrt{\alpha_{\tau_{s}}}}\Big)\notag\\&+\sqrt{1-\alpha_{\tau_{s-1}}-\sigma_{\tau_{s}}^{2}}\boldsymbol{\epsilon}_{\theta}(\mathbf{x}^{\tau_{s}},\tau_{s})+\sigma_{\tau_{s}}\boldsymbol{\epsilon}.
\end{align}
With a smaller number of steps $S$ compared to the original $T$, the reverse process can be accelerated.

\subsection {Thompson Sampling}
\label{TS}
Thompson sampling \citep{Thompson} has emerged as a prominent exploration strategy for decision-making under uncertainty \citep{bandit, t-uncertain}. To achieve a balance between exploration and exploitation \citep{balancing}, Thompson sampling utilizes a probability model to sample greedily based on the values of execution results from the last round. Specifically, given the value $v$, the probability model of Thompson sampling can be parameterized as $F(v,p)$, where $p$ is a random variable that ranges from $0$ to $1$. A higher value of $v$ can result in a higher sampling probability $\hat{p}$ from the probability model. Formally, at each round, we have the sampling probability:
\begin{gather}
\hat{p}\sim F(v,p).
\end{gather}
Then, Thompson sampling executes based on the sampling probability $\hat{p}$, updating the value $v$ and probability model $F(v,p)$ based on the execution results of the last round.


\subsection{Task Formulation}

For generative sequential recommendation, the goal is to generate the next item tailored to the target user conditioned on their historical interaction sequence.
The mainstream solutions to this task are from the embedding perspective.
Formally, we denote a user's historical interaction sequence as $\mathbf{e}_{1:N-1}=[\mathbf{e}_1,\mathbf{e}_2,\ldots,\mathbf{e}_{N-1}]$, where $\mathbf{e}_n$ represents the embedding of the $n$-th item the user has interacted with in chronological order. The subsequent item of this sequence, which we aim to generate, is represented as $\mathbf{e}_N$.
To apply diffusion models in generative recommendation, following prior studies \citep{DreamRec,Diffris}, noise is first added to $\mathbf{e}_N^0$ (equivalent to $\mathbf{e}_N$), followed by a denoising process leveraging the guidance signal $\mathbf{g}$ extracted from interaction history $\mathbf{e}_{1:N-1}$ to ensure the generated oracle items align closely with user preferences. The core is to model the item generation distribution $p_\theta(\mathbf{e}_N^{t-1} |\mathbf{e}_N^t, \mathbf{g})$ at each $t$-th denoising step, and inference step by step to generate the oracle items $\mathbf{e}_N^0$.

%% file: 3_methodology.tex
\section{Methodology}

\begin{figure*}
  \centering
  \includegraphics[width=1\textwidth]{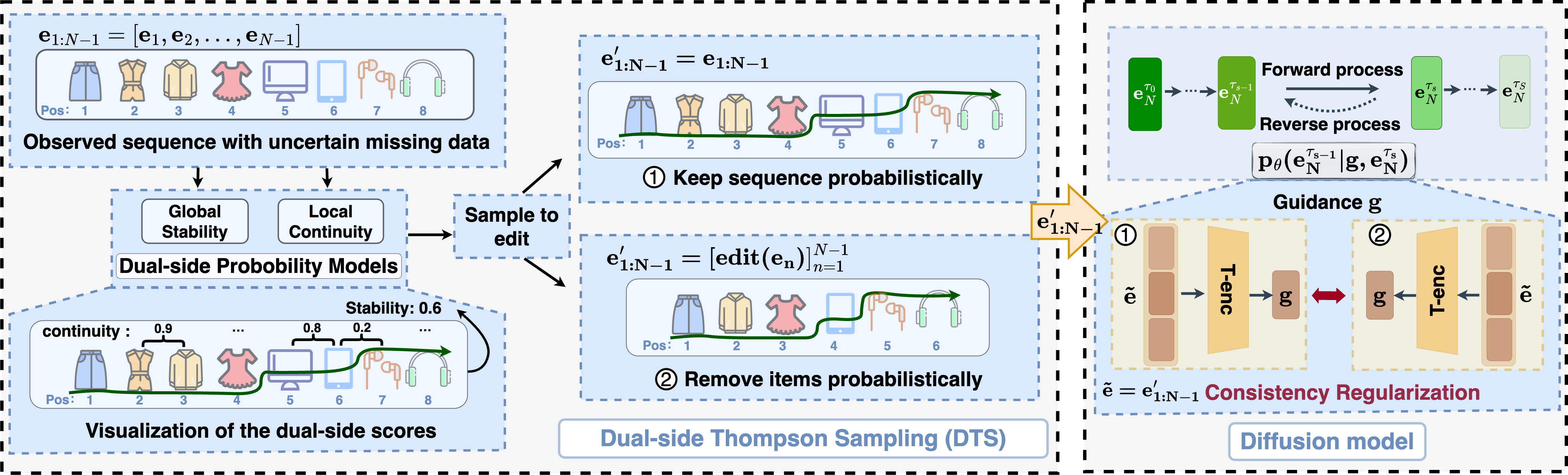}
   \vspace{-6pt}
  \caption{The overview of the TDM framework, which simulates extra missing data with DTS in the guidance signals, achieving diffusion models' consistency regularization and extrapolating to address the existing missing data. }
   \vspace{-10pt}
  \label{fig: method}
\end{figure*}

In this section, we present our proposed TDM, designed to mitigate the impact of missing data, as shown in Figure \ref{fig: method}.  
We begin by detailing the dual-side Thompson sampling (DTS) strategy in Section \ref{guidance}, which can simulate the mechanisms of missing data in user behaviors. 
Next, we describe the learning and generating phases of TDM in Section \ref{diffusion}. Finally, we provide theoretical analysis for TDM in Section \ref{theory}.

\subsection{Dual-side Thompson Sampling}
\label{guidance}
In real-world scenarios, missing data is inherently uncertain \citep{uncertainty1,uncertainty2} and hard to recover, so we simulate extra missing data in diffusion models' guidance signals, extrapolating to address the existing missing data. To preserve user preference evolution in sequences during simulation, we introduce a DTS strategy, sampling and removing items by exploiting user preference evolution with two probability models --- one at the local item level and the other at the global sequence level.

\subsubsection{Definition of Two Probability Models}
\label{probability model}
User preferences often exhibit dynamic shifts between items within an interaction sequence. To capture these preference shifts locally, we introduce the concept of \textit{continuity scores} to measure the similarity between adjacent items.
Formally, given a sequence $\mathbf{e}_{1:N-1}=[\mathbf{e}_1,\mathbf{e}_2,\ldots,\mathbf{e}_{N-1}]$, the continuity score for each item $\mathbf{e}_n$ within the sequence is:
\begin{gather}
\label{equ:un}
\mathrm{con}_n=\frac{\mathrm{exp}(\mathrm{sim}(\mathbf{e}_n,\mathbf{e}_{n+1}))}{\sum_{n'=1}^{N-2}\mathrm{exp}(\mathrm{sim}(\mathbf{e}_{n'},\mathbf{e}_{n'+1}))}, \quad n=1,2,\ldots, N-2,
\end{gather}
where $\mathrm{sim}(\cdot,\cdot)$ represents the cosine similarity function, $\mathrm{con}_n$ is the continuity score normalized using the \textit{softmax} function.
Intuitively, a higher continuity score indicates a greater similarity between adjacent items, indicating a stronger level of shared preference. 

In addition to local preference shifts, user preferences often fluctuate throughout the entire sequence \citep{fluctuation}. To assess the degree of these preference fluctuations, {we calculate the entropy value $h$ for each sequence with the continuity scores within it as below:
\begin{gather}
    h=-\sum_{n=1}^{N-2} \mathrm{con}_n\log (\mathrm{con}_n).
\end{gather}
Then, we define the \textit{stability score} $\mathrm{sta}_k$ for each sequence by normalizing its entropy value $h_{k}$ with softmax in a batch.
\begin{gather}
\mathrm{sta}_k=\frac{\mathrm{exp}(h_k)}{\sum_{k'=1}^{K}\mathrm{exp}(h_{k'})}, \quad k=1,2,\ldots, K,
\label{equ:tk}
\end{gather}
where $K$ is the number of sequences in the batch.}
A higher stability score reflects a higher entropy value,
suggesting that user preferences remain largely unchanged throughout the sequence.

We then parameterize the two probability models with value $\mathrm{con}_n$ and $\mathrm{sta}_k$ as introduced in Section \ref{TS}. Formally, we define the local item-side probability model as $L(\mathrm{con}_n,p_n)$ and the global sequence-side model as $G(\mathrm{sta}_k,p_k)$, where $p_n$ and $p_k$ are the random variables ranging from $0$ to $1$. Higher values in $\mathrm{con}_n$ and $\mathrm{sta}_k$ lead to higher sampling probabilities from their respective models.

\subsubsection{Strategical Editing with DTS}
To simulate missing data while maintaining user preference evolution patterns in sequences, the DTS samples sequences to edit and items to remove based on the two probability models defined in Section \ref{probability model}. 
Formally, we have:
\begin{gather}
\hat{p}_{n}\sim L(\mathrm{con}_n,p_n),\quad \hat{p}_{k}\sim G(\mathrm{sta}_k,p_k),
\end{gather}
where $\hat{p}_{k}$ is the sampling probabilities of the $k$-th sequence $\mathbf{e}_{1:N-1}$ to be edited, and $\hat{p}_{n}$ is probability for the $n$-th item $\mathbf{e}_n (1 \le n \le N-2)$ within sequence $\mathbf{e}_{1:N-1}$ to be discarded. To preserve the last item $\mathbf{e}_{N-1}$ in the sequence, we manually set $\hat{p}_{N-1}=0$.
Since $\mathrm{con}_n$ and $\mathrm{sta}_k$ represent the local continuity and global stability respectively, the dual-side Thompson sampling strategy tends to sample items with higher continuity in sequences with greater stability scores to remove. 
Removing such data is expected to have little impact on the original preference shifts, as shown in the consistent green curves in Figure \ref{fig: method}.
Therefore, we can simulate uncertain data missing, while preserving the underlying preference evolution pattern.

Having established $\hat{p}_{k}$ and $\hat{p}_{n}$, we can decide whether the $n$-th item in the $k$-th sequence would be discarded.
Formally, the strategically edited sequence is obtained as:
\begin{gather}
\label{equ:edit}
\mathbf{e}_{1:N-1}'=\begin{cases}\left[\,\text{edit}(\mathbf{e}_n)\,\right]_{n=1}^{N-1}  &\text{ if }1-\hat{p}_{k} < \lambda_1 \\
{\mathbf{e}_{1:N-1}}&\text{otherwise}\end{cases},
\\
\label{equ:remove}
\text{edit}(\mathbf{e}_n)=\begin{cases}\Phi  &\text{ if }1-\hat{p}_{n} < \lambda_2 \\
{\mathbf{e}_n}&\text{otherwise}\end{cases},
\end{gather}
where $\Phi$ is a dummy token, $\lambda_1$ and $\lambda_2$ are thresholds for sampling probabilities $\hat{p}_{k}$ and $\hat{p}_{n}$, ranging from $0$ to $1$, which control the proportion of removed items.
Higher values of $\lambda_1$ and $\lambda_2$ result in more items in more sequences being removed.

We then encode the strategically edited sequence $\mathbf{e}_{1:N-1}'$ as the guidance signals $\mathbf{g}$ using a Transformer encoder \text{T-enc}:
\begin{gather}
\mathbf{g}=\text{T-enc}(\mathbf{e}_{1:N-1}').
\end{gather}
In this way, the guidance is established after simulating extra missing data while preserving the evolution of users' dynamic preferences with the DTS strategy.

\subsection{Diffusion Model for Recommendation}
\label{diffusion}
Having acquired the guidance $\mathbf{g}$ as described in Section \ref{guidance}, we then leverage $\mathbf{g}$ to guide diffusion models to denoise, enabling TDM to recommend items robustly in the presence of missing data. 
To accelerate the generation during inference, we employ denoising diffusion implicit models introduced in Section \ref{ddim} for TDM to generate oracle items. 
Below, we detail TDM's training and generating phases.

\subsubsection{Training Phase}
\label{training}
For joint training of both conditional and unconditional models, we train TDM under the classifier-free guidance paradigm \citep{classifier-free}. 
Specifically, we randomly replace the guidance $\mathbf{g}$ with a dummy token $\Phi$ with probability $\rho$, while keeping the others unchanged. 
We view the next item $\mathbf{e}_{N}$ as the input $\mathbf{e}_{N}^0$ and add noise to it, following: $\mathbf{e}_{N}^{\tau_s}=\sqrt{\alpha_{\tau_s}}\mathbf{e}_N^0+\sqrt{1-\alpha_{\tau_s}}\boldsymbol{\epsilon}$.
Similar to DreamRec \citep{DreamRec}, we employ an MLP as the denoising neural network $f_\theta(\cdot,\cdot,\cdot)$ to directly predict $\mathbf{e}_{N}^{\tau_s}$ into $\hat{\mathbf{e}}_{N}^0$, rather than the noise $\boldsymbol{\epsilon}$, guided by $\mathbf{g}$:
\begin{gather}
\hat{\mathbf{e}}_{N}^0=f_\theta(\sqrt{\alpha_{\tau_s}}\mathbf{e}_N^0+\sqrt{1-\alpha_{\tau_s}}\boldsymbol{\epsilon},\mathbf{g},\tau_s),
\end{gather}
where $\hat{\mathbf{e}}_{N}^0$ denotes the prediction of $\mathbf{e}_{N}^{0}$.
According to Equation \eqref{loss_pre}, the loss function of TDM can be formulated as:
\begin{gather}
\label{loss_ddim}
\mathcal{L}=\sum_{s=1}^S\frac1{2d\sigma_{\tau_s}^2(1-\alpha_{\tau_s})}\mathbb{E}_{\mathbf{e}_N^0,\boldsymbol{\epsilon}}\left[\left\|\hat{\mathbf{e}}_N^0-\mathbf{e}_N^0\right\|_2^2\right].
\end{gather}

\subsubsection{Generating Phase}
Having trained the denoising model $f_\theta(\cdot,\cdot,\cdot)$, TDM can generate the oracle items step by step. 
Specifically, to integrate the conditional and unconditional generation under the classifier-free guidance paradigm, the denoising function is modified with a linear combination: 
\begin{gather}
\tilde{f}_\theta(\mathbf{e}_N^{\tau_s},\mathbf{g},\tau_s)=(1+w) f_\theta(\mathbf{e}_N^{{\tau_s}},\mathbf{g},\tau_s)-w f_\theta(\mathbf{e}_N^{\tau_s},\Phi,\tau_s),
\end{gather}
where the hyperparameter $w$ controls the guidance strength. 
A high value of $w$ increases reliance on the guidance $\mathbf{g}$, but it may lead to overfitting. 
Following Equation \eqref{reverse}, the reverse denoising step from $\tau_s$ to $\tau_{s-1}$ can be expressed as:
\begin{equation}
\begin{aligned}
\label{eq:denoise_one_step}
\mathbf{e}_N^{\tau_{s-1}}=&\sqrt{\alpha_{\tau_{s-1}}}\tilde{f}_\theta(\mathbf{e}_N^{\tau_s},\mathbf{g},\tau_s)\\&+\sqrt{1-\alpha_{\tau_{s-1}}}\frac{\mathbf{e}_N^{\tau_s}-\sqrt{\alpha_{\tau_s}}\left.\tilde{f}_\theta(\mathbf{e}_N^{\tau_s},\mathbf{g},\tau_s)\right.}{\sqrt{1-\alpha_{\tau_s}}}.
\end{aligned}
\end{equation}
Under the guidance $\mathbf{g}$ encoded from a interaction sequence, the oracel item $\mathbf{e}_N^0$ is generated by denosing a Gaussian sample $\mathbf{e}_N^{\tau_S}\sim \mathcal{N}(\mathbf{0},\mathbf{I})$ for ${\tau_S}$ times with Equation \eqref{eq:denoise_one_step}.
Once the oracle item is generated, we retrieve the K-nearest items from the candidate set to provide the top-K recommendation results.
See \url{https://github.com/maowenyu-11/TDM} for the algorithms of the training and generating phases of TDM.

\subsection{Theoretical Analysis}
\label{theory}
{Here we justify TDM in simulating extra missing data, which can enable diffusion models to address existing missing data through extrapolation and consistency regularization.}

\vspace{5pt}
\noindent\textbf{Extrapolation:} Let $\hat{\mathbf{g}}$, $\bar{\mathbf{g}}$, and $\tilde{\mathbf{g}}$ denote the guidance encoded from observed sequence $\mathbf{e}_{1:N-1}$, unavailable complete sequence $\mathbf{e}_{1:N-1}\oplus \delta$ and simulated sequence $\mathbf{e}_{1:N-1}'=\mathbf{e}_{1:N-1}\ominus \delta'$, respectively, where $\delta$ denotes the real missing data and $\delta'$ denotes the simulated missing data. Our objective is to demonstrate the validity of the extrapolation, specifically the inequality $\|f_\theta(\mathbf{e}^{\tau_s}_N,\bar{\mathbf{g}},\tau_s)-f_\theta(\mathbf{e}^{\tau_s}_N,\hat{\mathbf{g}},\tau_s)\|\le C\|f_\theta(\mathbf{e}^{\tau_s}_N,\hat{\mathbf{g}},\tau_s)-f_\theta(\mathbf{e}^{\tau_s}_N,\tilde{\mathbf{g}},\tau_s)\|$ for some constant $C$, where $\mathbf{e}^{\tau_s}_N$ represents the next interacted item with noise of $\tau_s$ time steps, $f_\theta$ is the denoising model.

{
Applying Taylor's Formula, we can express the two functions as follows:
\begin{equation}
\begin{aligned}
f_\theta(\mathbf{e}^{\tau_s}_{N},\bar{\mathbf{g}},\tau_s)=&f_\theta(\mathbf{e}^{\tau_s}_{N},\hat{\mathbf{g}},\tau_s)+(\bar{\mathbf{g}}-\hat{\mathbf{g}})^\top\nabla_g f_\theta(\mathbf{e}^{\tau_s}_{N},\hat{\mathbf{g}},\tau_s)\\&+o(\|\bar{\mathbf{g}}-\hat{\mathbf{g}}\|),\\f_\theta(\mathbf{e}^{\tau_s}_{N},\tilde{\mathbf{g}},\tau_s)=&f_\theta(\mathbf{e}^{\tau_s}_{N},\hat{\mathbf{g}},\tau_s)+(\tilde{\mathbf{g}}-\hat{\mathbf{g}})^\top\nabla_g f_\theta(\mathbf{e}^{\tau_s}_{N},\hat{\mathbf{g}},\tau_s)\\&+o(\|\tilde{\mathbf{g}}-\hat{\mathbf{g}}\|).
\end{aligned}
\end{equation}
Then, combining the above two equalities, we have
\begin{align}&f_\theta(\mathbf{e}^{\tau_s}_{N},\bar{\mathbf{g}},\tau_s)-f_\theta(\mathbf{e}^{\tau_s}_{N},\hat{\mathbf{g}},\tau_s)\notag\\&=(\bar{\mathbf{g}}-\hat{\mathbf{g}})^\top\nabla f_\theta(\mathbf{e}^{\tau_s}_{N},\hat{\mathbf{g}},\tau_s)+o(\|\bar{\mathbf{g}}-\hat{\mathbf{g}}\|)\notag\\&=\frac{(\bar{\mathbf{g}}-\hat{\mathbf{g}})^\top\nabla_g f_\theta(\mathbf{e}^{\tau_s}_{N},\hat{\mathbf{g}},\tau_s)}{(\tilde{\mathbf{g}}-\hat{\mathbf{g}})^\top\nabla_g f_\theta(\mathbf{e}^{\tau_s}_{N},\hat{\mathbf{g}},\tau_s)}(\tilde{\mathbf{g}}-\hat{\mathbf{g}})^\top\nabla_g f_\theta(\mathbf{e}^{\tau_s}_{N},\hat{\mathbf{g}},\tau_s)+o(\|\bar{\mathbf{g}}-\hat{\mathbf{g}}\|)\notag\\&=\frac{(\bar{\mathbf{g}}-\hat{\mathbf{g}})^\top\nabla_g f_\theta(\mathbf{e}^{\tau_s}_{N},\hat{\mathbf{g}},\tau_s)}{(\tilde{\mathbf{g}}-\hat{\mathbf{g}})^\top\nabla_g f_\theta(\mathbf{e}^{\tau_s}_{N},\hat{\mathbf{g}},\tau_s)}(f_\theta(\mathbf{e}^{\tau_s}_{N},\tilde{\mathbf{g}},\tau_s)-f_\theta(\mathbf{e}^{\tau_s}_{N},\hat{\mathbf{g}},\tau_s)) \notag \\& +o(\|\bar{\mathbf{g}}-\hat{\mathbf{g}}\|)+o(\|\tilde{\mathbf{g}}-\hat{\mathbf{g}}\|),
\end{align}
where we assume that $(\tilde{\mathbf{g}}-\hat{\mathbf{g}})^\top\nabla_g f_\theta(\mathbf{e}^{\tau_s}_{N},\hat{\mathbf{g}},\tau_s)\neq 0$. Thus, we obtain the following inequality:
\begin{align}&\|f_\theta(\mathbf{e}^{\tau_s}_{N},\bar{\mathbf{g}},\tau_s)-f_\theta(\mathbf{e}^{\tau_s}_{N},\hat{\mathbf{g}},\tau_s)\|\notag \\&\le\left|\frac{(\bar{\mathbf{g}}-\hat{\mathbf{g}})^\top\nabla_g f_\theta(\mathbf{e}^{\tau_s}_{N},\hat{\mathbf{g}},\tau_s)}{(\tilde{\mathbf{g}}-\hat{\mathbf{g}})^\top\nabla_g f_\theta(\mathbf{e}^{\tau_s}_{N},\hat{\mathbf{g}},\tau_s)}\right|\|f_\theta(\mathbf{e}^{\tau_s}_{N},\hat{\mathbf{g}},\tau_s)-f_\theta(\mathbf{e}^{\tau_s}_{N},\tilde{\mathbf{g}},\tau_s)\| \notag \\& +o(\|\bar{\mathbf{g}}-\hat{\mathbf{g}}\|)+o(\|\tilde{\mathbf{g}}-\hat{\mathbf{g}}\|),
\end{align}
where $\|f_\theta(\mathbf{e}^{\tau_s}_{N},\hat{\mathbf{g}},\tau_s)-f_\theta(\mathbf{e}^{\tau_s}_{N},\tilde{\mathbf{g}},\tau_s)\|$ is the distance between the prediction from observed sequence and simulated sequence. To bound the coefficients $\left|\frac{(\bar{\mathbf{g}}-\hat{\mathbf{g}})^\top\nabla_g f_\theta(\mathbf{e}^{\tau_s}_{N},\hat{\mathbf{g}},\tau_s)}{(\tilde{\mathbf{g}}-\hat{\mathbf{g}})^\top\nabla_g f_\theta(\mathbf{e}^{\tau_s}_{N},\hat{\mathbf{g}},\tau_s)}\right|$, we need to analyze the two inner products, ${(\bar{\mathbf{g}}-\hat{\mathbf{g}})^\top\nabla_g f_\theta(\mathbf{e}^{\tau_s}_{N},\hat{\mathbf{g}},\tau_s)}$ and ${(\tilde{\mathbf{g}}-\hat{\mathbf{g}})^\top\nabla_g f_\theta(\mathbf{e}^{\tau_s}_{N},\hat{\mathbf{g}},\tau_s)}$. 

If we can \textbf{simulate the mechanism of missing data} --- specifically, if the missing data process from $\mathbf{e}_{1:N-1}$ to $\mathbf{e}_{1:N-1}\ominus \delta'$ can align well with that from $\mathbf{e}_{1:N-1}\oplus\delta$ to $\mathbf{e}_{1:N-1}$ --- the difference between the two pair of data will be roughly equivalent. Consequently, the two differences in guidance, $\bar{\mathbf{g}}-\hat{\mathbf{g}}$ and $\hat{\mathbf{g}}-\tilde{\mathbf{g}}$, will also be approximately equal. In this scenario, the coefficient will be close to 1, resulting in a bounded value $C>0$. 

Thus we can validate that enhancing diffusion models' insensitivity to simulated missing data enables resilience against real missing data. Here, we implement the simulation mechanism as Dual-side Thompson Sampling (DTS), which preserves user preferences in sequences throughout the simulation.
\label{theory:extrapolation}
}

{\vspace{5pt}
\noindent\textbf{Consistency regularization:} To ensure the effectiveness of extrapolation, we leverage consistency regularization to minimize $\|f_\theta(\mathbf{e}^{\tau_s}_{N},\hat{\mathbf{g}},\tau_s)-f_\theta(\mathbf{e}^{\tau_s}_{N},\tilde{\mathbf{g}},\tau_s)\|$. Since the interaction sequences are edited probabilistically across different epochs, $\hat{\mathbf{g}}$ and $\tilde{\mathbf{g}}$ can serve as the perturbated pairs. Let the ground-truth label of the next item be $\mathbf{e}_N^0$. 
By completing the square, we obtain the inequality: 
\begin{align}
&||f_\theta(\mathbf{e}_N^{\tau_s},\hat{\mathbf{g}}, \tau_{s})-f_\theta(\mathbf{e}_N^{\tau_s},\tilde{\mathbf{g}}, \tau_{s})||_2^2 \notag \\& \leq  2\left(||f_\theta(\mathbf{e}_N^{\tau_s},\hat{\mathbf{g}},\tau_{s})\mathbf{e}_N^0||_2^2+||f_\theta(\mathbf{e}_N^{\tau_s},\tilde{\mathbf{g}}, \tau_{s})-\mathbf{e}_N^0||_2^2\right).
\end{align}
Consequently, we can achieve consistency regularization of minimizing the left-hand side by minimizing the right-hand side, which stems from our reconstruction loss. Such consistency regularization endows diffusion models with insensitivity to the simulated missing data, allowing the extrapolation to resist the real missing data issue. 
}

%% file: 4_experiments.tex
\section{Experiments}
In this section, we conduct extensive experiments across three datasets to evaluate the effectiveness of TDM
by answering the following questions.
\textbf{RQ1:} How does TDM perform in the sequential recommendation compared with diverse baseline models? 
\textbf{RQ2:} What are the respective contributions of probability models and denoising diffusion implicit models to our method?
\textbf{RQ3:} How sensitive is TDM to the thresholds of removing?
\textbf{RQ4:} How robust is TDM to varying degrees of missing data in datasets and different sequence lengths?
\textbf{RQ5:} Can DTS generalize on traditional recommender systems rather than diffusion models?

\subsection{Experimental Settings}
\vspace{5pt}
\noindent
\textbf{Datasets.}
We conduct experiments on three real-world datasets for sequential recommendation following the settings of DreamRec~\citep{DreamRec}: YooChoose~\citep{YooChoose},
KuaiRec~\citep{KuaiRec}, and Zhihu~\citep{zhihu}. To mitigate cold-start issues,
we implement a preprocessing step that excludes items with fewer than five interactions and sequences shorter than 3 interactions.
For each dataset, we sort all sequences chronologically and split the data into training, validation, and testing sets in an 8:1:1 ratio, ensuring that later interactions don't leak into the training data \citep{critic}. Additionally, to validate the effectiveness of TDM on larger or diverse datasets from different domains, we also conduct experiments on Steam, Amazon-beauty, and Amazon-toys. The detailed statistics of datasets are
provided in Table \ref{table:dataset}.

\begin{table}[t]
\renewcommand\arraystretch{1.05}
\caption{Statistics of the five datasets. }
\vspace{-2mm}
\label{table:dataset}
\centering
\setlength{\tabcolsep}{0.5mm}
\small
\begin{tabular}{ccccccc}
\toprule
Dataset& YooChoose& KuaiRec& Zhihu& Steam& Beauty& Toys
 \\
 \midrule
 \#sequences &128,468& 92,090& 11,714&281,428 & 22,363& 19,4124\\
\#items& 9,514 &7,261& 4,838& 13,044 &12,101 & 11,924\\
\#interactions& 539,436& 737,163 &77,712& 3,485,022& 198,502  &167,597\\
\toprule
\vspace{-6mm}
\end{tabular}
\end{table}

\vspace{5pt}
\noindent\textbf{Baselines.}
We compare the performance of TDM against multiple leading approaches, including:
\begin{itemize}[leftmargin=*]
\item Traditional sequential recommenders: GRU4Rec \citep{session},
Caser \citep{caser}, SASRec \citep{SASRec}, Bert4Rec \citep{Bert4Rec}, CL4SRec \citep{CL4Rec}, IPS \citep{IPS-S} and AdaRanker \citep{AdaRank}, which employ neural networks to model data distribution and capture user preferences. 

\item Generative recommenders: DiffRec \citep{DiffRec}, DiffRIS \citep{Diffris}, and DreamRec \citep{DreamRec}, which generate target items to recommend directly with diffusion models.

\item Recovering-based algorithms: DiffuASR \citep{DiffuASR}, CaDiRec \citep{cadirec}, PDRec \citep{PDRec}, {STEAM \citep{STEAM}, and SSDRec \citep{SSDRec}}. DiffASR, CaDiRec, and PDRec generate supplement items to the observed sequences with diffusion models to enhance traditional sequential recommenders. STEAM and SSDRec aim to correct the interaction sequence by ``insert'' or other operations.
\end{itemize}

\vspace{5pt}
\noindent\textbf{Implementation Details.}
Following DreamRec \citep{DreamRec}, we set the sequence length to 10, padding sequences with fewer than 10 interactions using a padding token. 
Our experiments are implemented using Python 3.9 and PyTorch 2.0.1, with computations performed on Nvidia GeForce RTX 3090 GPUs. The dimension of item embeddings is 64 across all models. The learning rate is tuned within the range of $[0.01, 0.005, 0.001, 0.0005, 0.0001, 0.00005]$.
For diffusion models, we varied the total diffusion step $T$ across
$[500,1000,2000]$, employing intervals of $100$ to obtain corresponding $\tau_S$ values of $[5,10,20]$ for denoising diffusion implicit models. The guidance strength $w$ is set within the range $[0, 2, 4, 6,8, 10]$, and the threshold $\lambda_1,\lambda_2$ are tuning across the range $[0,0.1, \ldots,1]$. We set the unconditional training probability $\rho$ as $0.1$ suggested by \citet{p-uncon}. We adopt the widely used metrics in sequential recommendation: hit ratio (HR@20) and normalized discounted cumulative gain (NDCG@20) \citep{SASRec} to evaluate the recommendation performance.
In our result tables, we report the average performance of five experimental runs, with their corresponding standard deviations. 

\subsection{Main Results (RQ1)}
To answer RQ1, we compare the recommendation performance of TDM against multiple baselines.
Table \ref{table:main_all} and Table \ref{table:main_add} present the experimental results on different datasets, demonstrating the superiority of TDM.
For example, on the KuaiRec dataset, TDM outperforms DreamRec, a generative recommender that utilizes diffusion models guided by observed sequences, with increases of $5.84\%$ and $13.84\%$ in HR@20 and NDCG@20, respectively. Similarly, on the YooChoose dataset, TDM outperforms PDRec, a recovering-based method, with improvements of $9.85\%$ in HR@20 and $26.95\%$ in NDCG@20.
These indicate that simulating missing data with DTS enhances the robustness of diffusion models to unreliable sequences, thereby improving overall recommendation performance.

\begin{table}[t]
\renewcommand\arraystretch{1.05}
\caption{Overall performance of different methods for the sequential recommendation. The best score and the second-best score are bolded and underlined, respectively. The last row indicates the performance improvements of TDM over the best-performing baseline method. }
\vspace{-2mm}
\label{table:main_all}
\centering
\setlength{\tabcolsep}{0.5mm}
\small
\begin{tabular}{ccccccc}
\toprule
\multirow{2}*{Methods} & \multicolumn{2}{c}{YooChoose}  & \multicolumn{2}{c}{KuaiRec} & \multicolumn{2}{c}{Zhihu} \\
\cmidrule(lr){2-3}\cmidrule(lr){4-5}\cmidrule(lr){6-7}
& HR(\%) & NDCG(\%) & HR(\%) & NDCG(\%) & HR(\%) & NDCG(\%)\\
\toprule
GRU4Rec & $3.89 \scriptstyle {\pm 0.11}$ & $1.62 \scriptstyle {\pm 0.02}$ & $3.32 \scriptstyle{\pm 0.11}$ & $1.23 \scriptstyle{\pm 0.08}$ & $1.78 \scriptstyle{\pm 0.12}$ & $0.67 \scriptstyle{\pm 0.03}$ \\
Caser & $4.06 \scriptstyle {\pm 0.12}$ & $1.88 \scriptstyle {\pm 0.09}$ & $2.88 \scriptstyle {\pm 0.19}$ & $1.07 \scriptstyle{\pm 0.07}$ & $1.57 \scriptstyle {\pm 0.05}$ & $0.59 \scriptstyle {\pm 0.01}$ \\
SASRec & $3.68 \scriptstyle {\pm 0.08}$ & $1.63 \scriptstyle {\pm 0.02}$ & $3.92 \scriptstyle {\pm 0.18}$ & $1.53 \scriptstyle {\pm 0.11}$ & $1.62 \scriptstyle {\pm 0.01}$ & $0.60 \scriptstyle {\pm 0.03}$ \\
Bert4Rec&$4.96\scriptstyle {\pm 0.05}$	&$2.05\scriptstyle {\pm 0.03}$&	$3.77\scriptstyle {\pm 0.09}$&$1.73\scriptstyle {\pm 0.04}$&$2.01\scriptstyle {\pm 0.06}$	&$0.72\scriptstyle {\pm 0.04}$\\
CL4SRec & $4.45 \scriptstyle {\pm 0.04}$ & $1.86 \scriptstyle {\pm 0.02}$ & $4.25 \scriptstyle {\pm 0.10}$ & $2.01 \scriptstyle {\pm 0.09}$ & $2.03 \scriptstyle {\pm 0.06}$ &$ 0.74 \scriptstyle {\pm 0.03}$ \\
IPS & $3.81 \scriptstyle{\pm 0.05}$ & $1.73 \scriptstyle {\pm 0.03}$ & $3.73 \scriptstyle {\pm 0.03}$ & $1.40 \scriptstyle {\pm 0.05}$ & $1.66 \scriptstyle {\pm 0.04}$ & $0.64 \scriptstyle {\pm 0.02}$ \\
AdaRanker & $3.74 \scriptstyle {\pm 0.06}$ & $1.67 \scriptstyle {\pm 0.04}$ & $4.14 \scriptstyle {\pm 0.09}$ & $1.89 \scriptstyle {\pm 0.05}$ & $1.70 \scriptstyle {\pm 0.04}$ & $0.61 \scriptstyle {\pm 0.02}$ \\
\midrule
  {STEAM}&  {$4.69 \scriptstyle{\pm 0.06}$} & {$1.76\scriptstyle{\pm0.02}$} &  {$4.98\scriptstyle{\pm 0.05}$} &  {$2.90 \scriptstyle{\pm 0.02}$} & {$1.75\scriptstyle{\pm 0.02}$} & {$0.69 \scriptstyle{\pm 0.02}$} \\
 {SSDRec}&  {$4.52 \scriptstyle{\pm 0.07}$} & {$1.95\scriptstyle{\pm0.03}$} & {$4.19 \scriptstyle{\pm 0.08}$} &  {$3.28 \scriptstyle{\pm 0.06}$} &  {$2.03 \scriptstyle{\pm 0.06}$} &  {$0.72 \scriptstyle{\pm 0.03}$} \\
DiffuASR&$4.48\scriptstyle {\pm0.03}$&$1.92\scriptstyle {\pm0.02}$&$4.53\scriptstyle {\pm 0.02}$&$3.30\scriptstyle {\pm0.03}$&$2.05\scriptstyle {\pm0.02}$&$0.71\scriptstyle {\pm0.02}$\\
CaDiRec &$5.05\scriptstyle {\pm0.05}$&$2.21\scriptstyle {\pm0.10}$&$2.56\scriptstyle {\pm 0.04}$&$1.79\scriptstyle {\pm 0.03}$&	$2.14\scriptstyle {\pm0.05}$&$0.72\scriptstyle {\pm0.07}$\\
PDRec&$\underline{6.22}\scriptstyle {\pm0.03}$ &$\underline{3.17}\scriptstyle {\pm0.02}$&$4.42\scriptstyle {\pm0.03}$& $ 3.55\scriptstyle {\pm0.04 }$&$2.10\scriptstyle {\pm0.03}$&$0.74\scriptstyle {\pm0.02 }$\\
\midrule
DiffRec & $4.33 \scriptstyle {\pm 0.02}$ & $1.84 \scriptstyle {\pm 0.01}$ & $3.74 \scriptstyle {\pm 0.08}$ & $1.77 \scriptstyle {\pm 0.05}$ & $1.82 \scriptstyle {\pm 0.03}$ & $0.65 \scriptstyle {\pm 0.09}$ \\
DiffRIS&$4.51 \scriptstyle {\pm0.03}$& $1.95 \scriptstyle {\pm0.02}$&$4.28 \scriptstyle {\pm0.03}$& $2.03 \scriptstyle {\pm0.04}$&-&-\\ 
DreamRec & $4.78 \scriptstyle {\pm 0.06}$ & $2.23 \scriptstyle {\pm 0.02}$ & $\underline{5.16} \scriptstyle {\pm 0.05}$ & $\underline{4.11} \scriptstyle {\pm 0.02}$ & $\underline{2.26} \scriptstyle {\pm 0.07}$ & $\underline{0.79} \scriptstyle {\pm 0.01}$  \\
\midrule
TDM &$\mathbf{6.90}\scriptstyle {\pm0.01}$&$\mathbf{4.34}\scriptstyle {\pm0.03}$
&$\mathbf{5.48}\scriptstyle {\pm0.02}$&$\mathbf{4.77}\scriptstyle {\pm0.04}$
&$\mathbf{2.65}\scriptstyle {\pm0.03}$&$\mathbf{0.88}\scriptstyle {\pm 0.04}$\\
\rowcolor{mygray} Improv.&$9.85\%$&$26.95\%$&$5.84\%$&$13.84\%$&$14.72\%$&$10.23\%$\\
\bottomrule
\vspace{-6mm}
\end{tabular}
\end{table}

\begin{table}[t]
\renewcommand\arraystretch{1.1}
\caption{Experimental results on larger dataset (Steam) and diverse datasets from different domains (Beauty and Toys).
}
\vspace{-2mm}
\label{table:main_add}
\centering
\setlength{\tabcolsep}{0.5mm}
\small
\begin{tabular}{ccccccc}
\toprule
\multirow{2}*{Methods} & \multicolumn{2}{c}{Steam}  & \multicolumn{2}{c}{Toys} & \multicolumn{2}{c}{Beauty} \\
\cmidrule(lr){2-3}\cmidrule(lr){4-5}\cmidrule(lr){6-7}
& HR(\%) & NDCG(\%) & HR(\%) & NDCG(\%) & HR(\%) & NDCG(\%) \\
\toprule
GRU4Rec  &$9.23 \scriptstyle {\pm 0.05}$ & $3.56 \scriptstyle {\pm 0.03}$ & $3.18 \scriptstyle {\pm 0.08}$ & $1.27 \scriptstyle {\pm 0.03}$ &$3.85 \scriptstyle {\pm 0.09}$ & $1.38 \scriptstyle {\pm 0.06}$  \\
 Caser  &$15.20 \scriptstyle {\pm 0.09}$ & $6.62 \scriptstyle {\pm 0.05}$ &$8.83 \scriptstyle {\pm 0.09}$  &$4.02 \scriptstyle {\pm 0.05}$ & $8.67 \scriptstyle {\pm 0.06}$ & $4.36 \scriptstyle {\pm 0.10}$  \\
 SASRec  & $13.61 \scriptstyle {\pm 0.06}$ &$5.36 \scriptstyle {\pm 0.08}$  &$9.23 \scriptstyle {\pm 0.07}$ & $4.33 \scriptstyle {\pm 0.02}$ & $8.98 \scriptstyle {\pm 0.12}$  & $3.66 \scriptstyle {\pm 0.07}$  \\
 Bert4Rec &$12.73 \scriptstyle {\pm 0.08}$&$5.20 \scriptstyle {\pm 0.07}$ & $4.59 \scriptstyle {\pm 0.08}$ &$1.90 \scriptstyle {\pm 0.06}$ &$5.79 \scriptstyle {\pm 0.11}$ &$2.35 \scriptstyle {\pm 0.12}$ \\
 CL4SRec & $15.06 \scriptstyle {\pm 0.08}$&$6.12 \scriptstyle {\pm 0.06}$&$9.09 \scriptstyle {\pm 0.03}$ & $5.08 \scriptstyle {\pm 0.03}$ &$10.18 \scriptstyle {\pm 0.11}$ & $4.85 \scriptstyle {\pm 0.12}$ \\
 IPS  &$15.65 \scriptstyle {\pm 0.08}$&$6.46 \scriptstyle {\pm 0.02}$ &$9.29 \scriptstyle {\pm 0.01}$&$5.27 \scriptstyle {\pm 0.04}$ &$10.15 \scriptstyle {\pm 0.02}$&$4.56 \scriptstyle {\pm 0.07}$ \\
 AdaRanker &$15.71 \scriptstyle {\pm 0.07}$ & $6.68 \scriptstyle {\pm 0.08}$ & $8.18 \scriptstyle {\pm 0.02}$ & $4.33 \scriptstyle {\pm 0.02}$ & $8.03 \scriptstyle {\pm 0.08}$&$3.80 \scriptstyle {\pm 0.06}$  \\
 \midrule
 DiffuASR & $15.74 \scriptstyle {\pm 0.04}$ & $6.59 \scriptstyle {\pm 0.06}$  &$\underline{9.39} \scriptstyle {\pm 0.04}$ & $5.19 \scriptstyle {\pm 0.06}$ & $10.03 \scriptstyle {\pm 0.06}$& $5.16 \scriptstyle {\pm 0.11}$  \\
 CaDiRe& $15.65 \scriptstyle {\pm 0.07}$ &$6.42 \scriptstyle {\pm 0.12}$  & $9.33 \scriptstyle {\pm 0.03}$  & $5.16 \scriptstyle {\pm 0.11}$ & $9.85 \scriptstyle {\pm 0.08}$  & $4.46 \scriptstyle {\pm 0.04}$  \\
 PDRec  & $\underline{15.78} \scriptstyle {\pm 0.07}$ & $6.51 \scriptstyle {\pm 0.08}$  & $9.08 \scriptstyle {\pm 0.08}$  & $5.12 \scriptstyle {\pm 0.06}$ & $10.24 \scriptstyle {\pm 0.06}$ &$\underline{5.02} \scriptstyle {\pm 0.09}$ \\
 \midrule
 DiffRec     & $15.09 \scriptstyle {\pm 0.04}$ & $\underline{6.89} \scriptstyle {\pm 0.03}$ &$9.18 \scriptstyle {\pm 0.06}$& $\underline{5.25} \scriptstyle {\pm 0.04}$ & $10.21 \scriptstyle {\pm 0.04}$ & $5.14 \scriptstyle {\pm 0.02}$ \\
 DreamRec  &$15.08 \scriptstyle {\pm 0.08}$ &$6.39 \scriptstyle {\pm 0.08}$ & $9.18 \scriptstyle {\pm 0.08}$  &$5.22 \scriptstyle {\pm 0.08}$&$\underline{10.32} \scriptstyle {\pm 0.03}$ & $4.88 \scriptstyle {\pm 0.07}$  \\
 \midrule
 TDM &$\mathbf{16.19} \scriptstyle {\pm 0.01}$&$\mathbf{7.52} \scriptstyle {\pm 0.03}$ & $\mathbf{9.88} \scriptstyle {\pm 0.01}$ & $\mathbf{5.39} \scriptstyle {\pm 0.03}$ & $\mathbf{10.72} \scriptstyle {\pm 0.06}$ &$\mathbf{5.40} \scriptstyle {\pm 0.04}$ \\
\rowcolor{mygray} 
Improv.&$2.53\%$ & $ 8.38\% $&$ 4.96\%$ & $2.60\% $ & $3.73\%$  &$7.04\%$   \\
\bottomrule
\vspace{-6mm}
\end{tabular}
\end{table}

\subsection{Ablation Study (RQ2)}
\label{sec:abla}
The two probability models are designed to capture the dynamic evolution of user preferences, thereby preserving these evolution patterns in sequences despite the extra missing data simulated by DTS. To evaluate the impacts of the two probability models, we conduct ablation studies with eight variants of TDM. The experimental results are shown in Table \ref{table:abla}. ``w/o L'' and ``w/o G'' indicate variants where the local or global probability model is replaced with random sampling, ``w/o GL'' denotes replacing both models with random sampling, ``Base'' represents generating oracle items without simulating missing data. To extend our method, we propose other metrics for probability models and compare their performance. Specifically, ``w/P'' and ``w/I'' denote parameterizing the local probability model with item \textbf{p}opularity score (measured by the frequency of each item in all interactions) or score of \textbf{i}tem position in the sequence. Meanwhile, ``w/D'' and ``w/S'' represent parameterizing the global probability model with intra-sequence \textbf{d}iversity score or score for \textbf{s}equence length.

As can be seen, almost all variants (\ie ``w/o GL'', ``w/o G'', ``w/o L'', ``w/P'', ``w/I'', ``w/D'', and ``w/S'') outperform the ``Base'' model, highlighting the effectiveness of simulating missing data in enhancing the robustness of diffusion models, and hence improving the recommendation performance.
Furthermore, our proposed TDM
outperforms other variants (including ``w/o L'', ``w/o G'', ``w/o GL'', ``w/P'', ``w/I'', ``w/D'', and ``w/S''). This demonstrates the effectiveness of continuity and stability metrics in accounting for the evolution of user preferences when generating extra missing data, which aligns more closely with the missing mechanisms discussed in the theoretical proof in Section \ref{theory:extrapolation}.

\begin{table}[t]
\renewcommand\arraystretch{1.1}
\caption{Ablation Study for the metrics of probability models.
}
\vspace{-2mm}
\label{table:abla}
\centering
\setlength{\tabcolsep}{0.5mm}
\small
\begin{tabular}{ccccccc}
\toprule
\multirow{2}*{Methods} & \multicolumn{2}{c}{YooChoose}  & \multicolumn{2}{c}{KuaiRec} & \multicolumn{2}{c}{Zhihu} \\
\cmidrule(lr){2-3}\cmidrule(lr){4-5}\cmidrule(lr){6-7}
& HR(\%) & NDCG(\%) & HR(\%) & NDCG(\%) & HR(\%) & NDCG(\%) \\
\toprule
Base & $4.78 \scriptstyle {\pm 0.06}$ & $2.23 \scriptstyle {\pm 0.02}$ & $5.16 \scriptstyle {\pm 0.05}$ & $4.11 \scriptstyle {\pm 0.02}$ & $2.26\scriptstyle {\pm 0.07}$ & $0.79 \scriptstyle {\pm 0.01}$ \\
w/o GL &$6.24\scriptstyle {\pm0.07}$&
$3.91\scriptstyle {\pm0.06}$&$5.37\scriptstyle {\pm0.05}$&
$4.19\scriptstyle {\pm0.06}$&$2.30\scriptstyle {\pm0.05}$&
$0.80\scriptstyle {\pm0.02 }$\\
w/o L&$6.41\scriptstyle {\pm0.06}$&$4.26\scriptstyle {\pm0.05}$
&$5.44\scriptstyle {\pm0.03}$&$4.63\scriptstyle {\pm0.02}$
&$2.34\scriptstyle {\pm0.03}$&$0.86\scriptstyle {\pm0.02}$\\
w/o G&$6.48\scriptstyle {\pm0.01}$	&$4.29\scriptstyle {\pm0.04}$
&$5.43\scriptstyle {\pm0.04}$&$4.64\scriptstyle {\pm0.02}$
&$2.44\scriptstyle {\pm0.02}$&$0.81\scriptstyle {\pm0.08}$
\\
w/ P&{$6.28 \scriptstyle{\pm 0.02}$ }& {$4.18\scriptstyle{\pm0.03}$} &{$5.46   \scriptstyle{\pm 0.05}$} & {$4.57 \scriptstyle{\pm 0.08}$} &{$2.38  \scriptstyle{\pm 0.03}$ }&{$0.80 \scriptstyle{\pm 0.07}$ }\\
w/ I& {$6.28  \scriptstyle {\pm 0.04}$} & {$3.96 \scriptstyle {\pm 0.02}$} & {$5.20  \scriptstyle {\pm 0.03}$}& {$4.55\scriptstyle {\pm0.04}$ }& {$2.24 \scriptstyle {\pm 0.08}$ }& {$0.79 \scriptstyle {\pm 0.05}$} \\
w/ D &{$6.26 \scriptstyle{\pm 0.04}$ }& {$4.20\scriptstyle{\pm0.06}$} & {$5.45 \scriptstyle{\pm 0.02}$} &{ $4.52 \scriptstyle{\pm 0.02}$} & {$2.29 \scriptstyle{\pm 0.02}$} & {$0.81 \scriptstyle{\pm 0.03}$ }\\
w/ S&{$6.27 \scriptstyle{\pm 0.03}$}& {$4.30\scriptstyle{\pm0.06}$} & {$5.46\scriptstyle{\pm 0.03}$} & {$4.54 \scriptstyle{\pm 0.08}$} &{$2.30  \scriptstyle{\pm 0.08}$} &{$0.83 \scriptstyle{\pm 0.06}$} \\
\midrule
TDM &$\mathbf{6.90}\scriptstyle {\pm0.01}$&$\mathbf{4.34}\scriptstyle {\pm0.03}$
&$\mathbf{5.48}\scriptstyle {\pm0.02}$&$\mathbf{4.77}\scriptstyle {\pm0.04}$
&$\mathbf{2.65}\scriptstyle {\pm0.03}$&$\mathbf{0.88}\scriptstyle {\pm 0.04}$\\
\toprule
\end{tabular}
\end{table}

Denoising diffusion implicit model (DDIM) \citep{DDIM} is designed to accelerate the reverse process while maintaining the comparable performance of denoising diffusion probabilistic model (DDPM) \citep{DDPM}. We conduct experiments to demonstrate the impact of denoising diffusion implicit models on the performance of diffusion-based recommenders. Specifically, we apply the two models to both DreamRec \citep{DreamRec} and TDM, yielding four variants. The experimental results are presented in Table \ref{table: DDIM/DDPM}.

\begin{table}[t]
\renewcommand\arraystretch{1.1}
\caption{
Performance comparison of different types of diffusion models. 
``-P'' denotes using DDPM \citep{DDPM}, while ``-I'' denotes using DDIM \citep{DDIM}.}
\vspace{-2mm}
\label{table: DDIM/DDPM}
\centering
\setlength{\tabcolsep}{0.5mm}
\small
\begin{tabular}{ccccccc}
\toprule
\multirow{2}*{Methods} & \multicolumn{2}{c}{YooChoose}  & \multicolumn{2}{c}{KuaiRec} & \multicolumn{2}{c}{Zhihu} \\
\cmidrule(lr){2-3}\cmidrule(lr){4-5}\cmidrule(lr){6-7}
& HR(\%) & NDCG(\%) & HR(\%) & NDCG(\%) & HR(\%) & NDCG(\%) \\
\toprule
DreamRec-P& $4.78 \scriptstyle{\pm 0.06}$  &$2.23\scriptstyle{\pm0.02}$ & $5.16 \scriptstyle{\pm 0.05}$ &$4.11 \scriptstyle{\pm 0.02}$ & $2.26 \scriptstyle{\pm 0.07}$&$0.79 \scriptstyle{\pm 0.01}$ \\
 DreamRec-I& $4.85    \scriptstyle{\pm 0.06}$ & $2.26\scriptstyle{\pm0.07}$ & $4.93 \scriptstyle{\pm 0.03}$ &$4.01 \scriptstyle{\pm 0.07}$ & $2.25 \scriptstyle{\pm 0.02}$ & $  0.82 \scriptstyle{\pm 0.02}$ \\
 TDM-P& $6.82 \scriptstyle{\pm 0.02}$ & $4.33\scriptstyle{\pm0.02}$ & $5.49 \scriptstyle{\pm 0.02}$ & $4.74 \scriptstyle{\pm 0.08}$ & $2.45 \scriptstyle{\pm 0.03}$ &$0.85 \scriptstyle{\pm 0.02}$ \\
 TDM-I & $6.90 \scriptstyle{\pm 0.01}$ &$4.34\scriptstyle{ \pm 0.03}$ & $5.48\scriptstyle{ \pm 0.02}$ &$4.77\scriptstyle{\pm 0.04}$ &$2.65\scriptstyle{ \pm 0.03}$ &$0.88 \scriptstyle{\pm 0.04}$ \\
\toprule
\vspace{-6mm}
\end{tabular}
\end{table}

As shown in Table \ref{table: DDIM/DDPM}, the performance of the two types of diffusion models on sequential recommendation tasks is comparable, as evidenced by similar results for DreamRec-P and DreamRec-I. Furthermore, TDM consistently outperforms DreamRec, demonstrating the effectiveness of our DTS in enhancing the robustness of diffusion models. This indicates that while DDIM accelerates the generation process, it does not improve recommendation performance. Instead, it is the DTS strategy that facilitates performance enhancement in diffusion-based sequential recommenders by addressing missing data issues.

\subsection{Sensitivity Analysis (RQ3)}

\begin{figure*}
  \centering
  \includegraphics[width=0.89\textwidth]{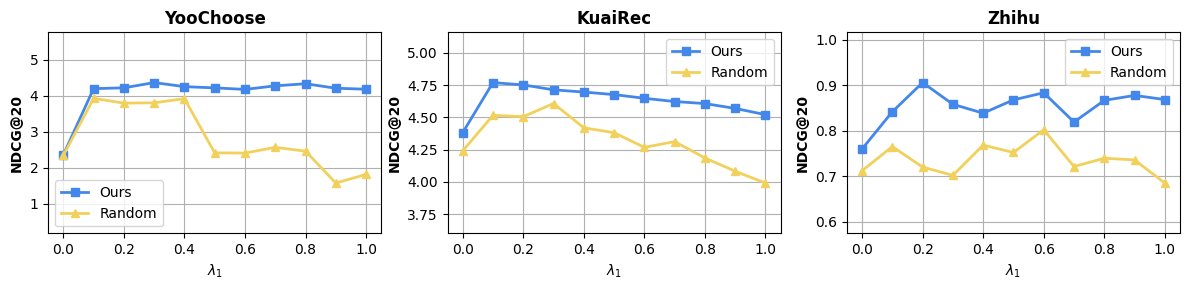}
   \vspace{-3mm}
  \caption{Sensitivity of TDM to the hyperparameter of $\lambda_1$ on multiple datasets, demonstrating the proportion of edited sequences. The ``random'' represents the variant ``w/o GL'' of TDM.}
   \vspace{-1mm}
  \label{fig: p1}
\end{figure*}
\begin{figure*}
  \centering
  \includegraphics[width=0.89\textwidth]{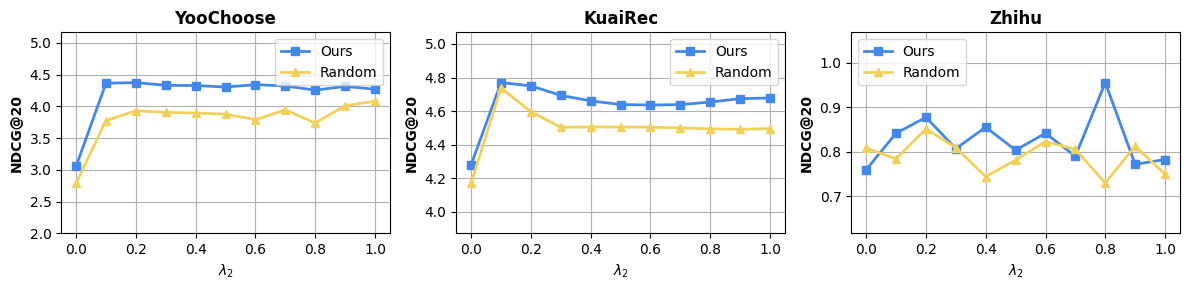}
   \vspace{-3mm}
  \caption{Sensitivity of TDM to the hyperparameter of $\lambda_2$ on multiple datasets, demonstrating the proportion of removed items. The ``random'' represents the variant ``w/o GL'' of TDM.}
   \vspace{-1mm}
  \label{fig: p2}
\end{figure*}

We examine the sensitivity of TDM to the parameters of threshold $\lambda_1$ and $\lambda_2$, which represent the proportion of edited sequences and removed items when simulating data missing in TDM. The experimental results are shown in Figure \ref{fig: p1} and \ref{fig: p2}. We can observe that the performance remains relatively stable; however, excessive or insufficient removal can result in suboptimal outcomes. This highlights the importance of choosing appropriate thresholds for DTS to simulate missing data.

\subsection{Robustness of TDM (RQ4)}
\label{synthetic}
\begin{figure*}
  \centering
  \includegraphics[width=0.95\textwidth]{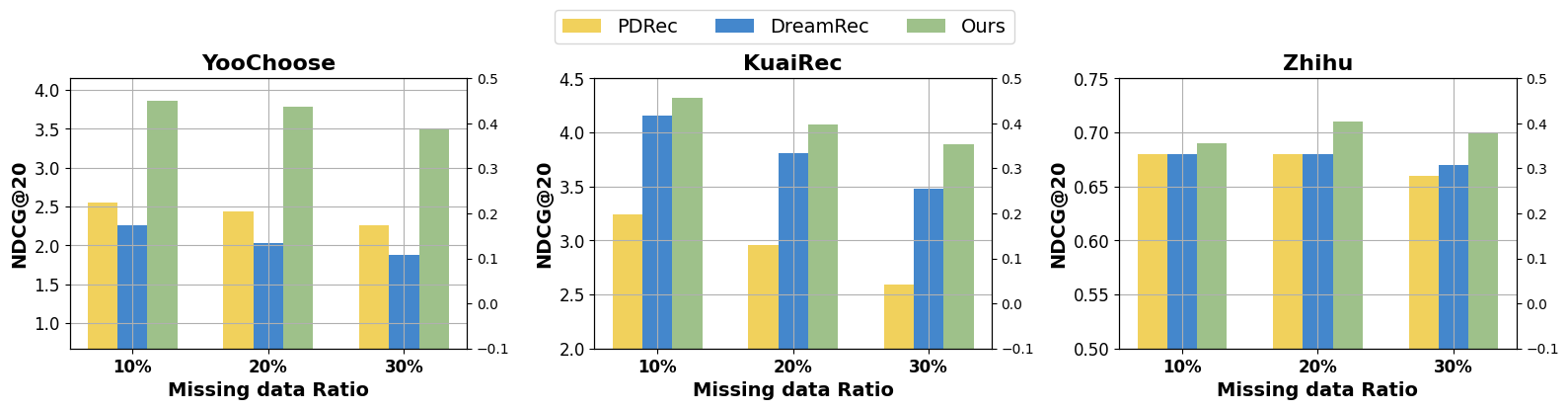}
   \vspace{-2mm}
  \caption{Performance of TDM on synthetic datasets with different missing ratios.}
  \label{fig: missing}
\end{figure*}

To validate the robustness of TDM to various missing data ratios, we create synthetic datasets with missing data proportions of $10\%$, $20\%$, and $30\%$ respectively. We compare the recommendation performance of TDM with representative baselines, including PDRec and DreamRec, across synthetic datasets. PDRec is a recovering-based method that utilizes diffusion models to enhance traditional recommenders, while DreamRec is a generative recommender where diffusion models generate items to recommend directly. As shown in Figure \ref{fig: missing}, TDM outperforms these baseline models on our synthetic datasets with varying missing ratios. Furthermore, the performance decline due to increased missing data in TDM is less significant than that observed in baseline models, as evidenced by the increasing height difference between the columns. These results demonstrate the superior robustness of TDM to varying degrees of missing data in datasets, as well as the effectiveness of DTS in addressing missing data in sequential recommendation.

\begin{figure*}
  \centering
  \includegraphics[width=0.95\textwidth]{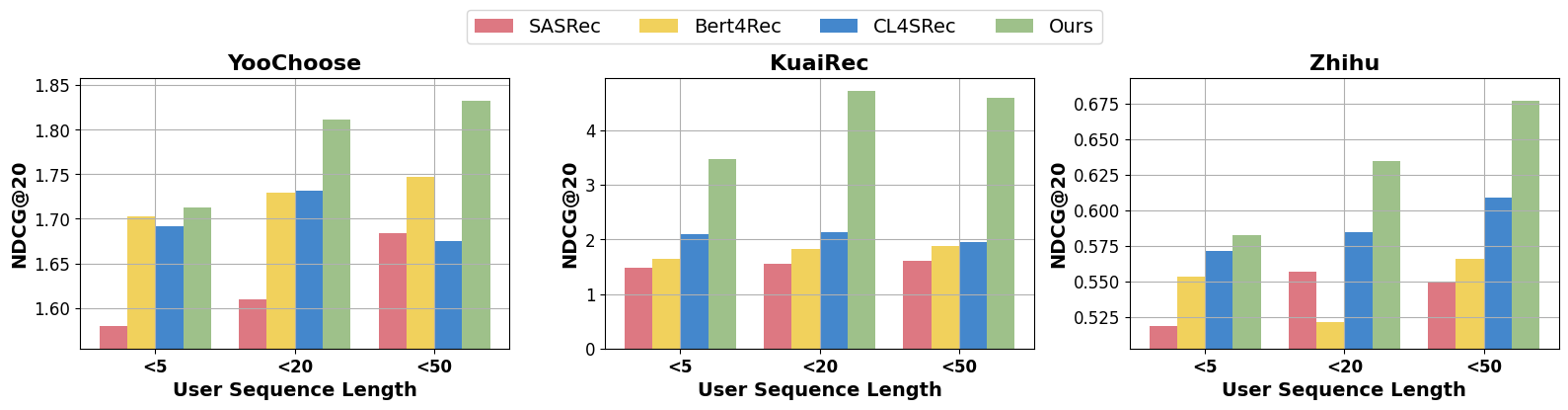}
   \vspace{-1mm}
  \caption{Performance of TDM on multiple datasets with different sequence lengths.}
    \vspace{-2mm}
  \label{fig: len}
\end{figure*}

We further investigate the robustness of TDM on datasets with different sequence lengths (\ie <5, <20, <50). We compare the recommendation performance of TDM with representative baseline models, including SASRec, Bert4Rec, and CL4SRec. The results are shown in Figure \ref{fig: len}. TDM maintains a significant lead on different sequence lengths, validating the robustness of TDM to different sequence lengths.  

\subsection{Generalization ability of TDM (RQ5)}
Introducing missing data with DTS can be considered as introducing a form of noise within the sequence, which can enhance diffusion models' denoising ability to missing data rather than Gaussian noise. 
However, DTS can be \textbf{a general algorithm to enhance recommendation systems' robustness} against uncertain missing data.
To validate the performance of DTS when extended to traditional recommenders, 
We conduct experiments on other recommenders, including GRU4Rec, SASRec,  CL4Rec, Caser, AdaRanker, and DiffuASR. The experimental results are presented in Table \ref{table:tr_gr}.

\begin{table}[t]
\renewcommand\arraystretch{1}
\caption{Comparison of TDM and traditional recommenders on performance improvement from the DTS strategy.}
\vspace{-2mm}
\label{table:tr_gr}
\centering
\setlength{\tabcolsep}{0.8mm}
\small
\begin{tabular}{ccccccc}
\toprule
\multirow{2}*{Methods} & \multicolumn{2}{c}{YooChoose}  & \multicolumn{2}{c}{KuaiRec} & \multicolumn{2}{c}{Zhihu} \\
\cmidrule(lr){2-3}\cmidrule(lr){4-5}\cmidrule(lr){6-7}
& HR(\%) & NDCG(\%) & HR(\%) & NDCG(\%) & HR(\%) & NDCG(\%) \\
\toprule
GRU4Rec & $3.89 $ & $1.62 $ & $3.32 $ & $1.23$ & $1.78$ & $0.67$ \\
+ DTS&$3.96$& $1.72$&$3.43$ &$1.40 $ & $1.83$& $0.68$ \\
\rowcolor{mygray} Improv.&$1.77\%$&$5.81\%$&$3.21\%$&$12.14\%$&$2.73\%$&$1.47\%$\\
\midrule
SASRec & $3.68$ & $1.63$ & $3.92$ & $1.53 $ & $1.62 $ & $0.60$ \\
+DTS&$3.98$&$1.58$ &$3.96$&$1.63$& $1.70$ & $0.72$\\
\rowcolor{mygray} Improv.&$7.54\%$&$-3.07\%$&$1.01\%$&$6.13\%$&$4.71\%$&$\mathbf{11.11}\%$\\
\midrule
CL4SRec & $4.45 $ & $1.86$ & $4.25$ & $2.01$ & $2.03$ &$ 0.74$ \\
+DTS & 
$4.63$& $1.88$&$4.57$& $2.25$ & $2.16$ & $0.78$ \\
\rowcolor{mygray} Improv.&$3.89\%$&$1.06\%$&$7.00\%$&$10.67\%$&$6.02\%$&$5.13\%$\\
\midrule
{
Caser}&{$ 4.06 $} &{$ 1.88 $ }&{$ 2.88 $ }&{$1.07 $} &{$ 1.57 $} &{$0.59$}\\
{+DTS} &{$ 4.27 $} &{$ 2.01 $ }&{$ 3.19$} &{$ 1.09$} &{$ 1.72 $ }&{$ 0.65$} \\
\rowcolor{mygray}
{Improv.} &{$ 4.92\% $} &{$ 6.47\% $} &{$ \mathbf{9.72}\%$} &{$ 1.83\%$} &{$ 8.72\% $} &{$9.23\% $}\\
\midrule
{AdaRanker} &{$ 3.74$ }&{$ 1.67 $} &{$ 4.14 $} &{$ 1.89 $ }&{$ 1.70 $ }&{$ 0.61 $} \\
{+DTS}  &{$ 4.16 $} &{$1.97$} &{$ 4.33 $} &{$1.93 $} &{$ 1.64 $} &{$ 0.67$ }\\
\rowcolor{mygray}
{Improv.} &{$10.10\% $} &{$ 15.22\% $} &{$ 4.39\% $} &{$2.07\% $ }&{$ -3.53\%$ }&{$ 8.96\% $} \\
\midrule
{DiffuASR}&{$4.48$} &{$ 1.92$ }&{$4.53$} &{$ 3.30 $} &{$ 2.05 $} &{$ 0.71$}\\
{+DTS }&{$ 4.66$}&{$2.08$ }&{$4.58 $ }&{$ 3.98 $} &{$ 2.05$} &{$0.73$}\\
\rowcolor{mygray}
{Improv.} &{$3.86\%$ }& {$7.69\%$} & {$1.09\%$} & {$\mathbf{17.09}\%$}&{$ 0.00\% $ }&{$ 2.82\%$}\\
\midrule
DreamRec & $4.78 $ & $2.23$ & $5.16$ & $4.11$ & $2.26$ & $0.79$ \\
+DTS &$6.90$&$4.34$
&$5.48$&$4.77$
&$2.65$&$0.88$\\
\rowcolor{mygray} Improv.&$\mathbf{30.72}\%$&$\mathbf{48.62}\%$&$5.84\%$&$13.84\%$&$\mathbf{14.72}\%$&$10.23\%$\\
\toprule
\vspace{-6mm}
\end{tabular}
\end{table}

As shown in Table \ref{table:tr_gr}, applying DTS to traditional recommenders can yield nearly universal improvement in recommendation performance.
This empirically proves the effectiveness and extensibility of DTS. 
Notably, when applied to DreamRec, a generative recommender using diffusion models, it yields the highest performance gains on average. This observation highlights that diffusion models provide a solid foundation for DTS to achieve consistency regularization from the empirical perspective, {owing to their capability to model complex data distributions and denoise missing data.}

\subsection{Computational Resource Comparison }

Since the Thompson sampling strategy relies solely on similarity computation and entropy calculation, its computational resource consumption is negligible compared to that of the transformer network architecture. As evidenced in Table \ref{table:time}, the computational complexity of TDM for training each epoch is nearly similar to other diffusion-based recommenders and traditional recommenders that use the same sequence encoder. Furthermore, by employing denoising diffusion implicit models to accelerate generation, we enhance the efficiency of TDM during the inference phase. As shown in Table \ref{table:time}, TDM substantially reduces the time cost during the inference phase than DreamRec and has a similar training time cost with other methods.

\begin{table}[htp]
\renewcommand\arraystretch{1.1}
\caption{Running time comparison of TDM and other methods on three datasets.}
\vspace{-2mm}
\label{table:time}
\centering
\small
\setlength{\tabcolsep}{0.5mm}
\begin{tabular}{ccccccc}
\toprule
\multirow{2}*{Methods} & \multicolumn{2}{c}{YooChoose}  & \multicolumn{2}{c}{KuaiRec} & \multicolumn{2}{c}{Zhihu} \\
\cmidrule(lr){2-3}\cmidrule(lr){4-5}\cmidrule(lr){6-7}
& Train & Inference & Train & Inference & Train & Inference 
 \\
 \midrule
{SASRec} &{01m 38s} &{00m 06s}&{02m 07s} &{00m 08s}&{00m 10s}& {00m 01s}\\
{AdaRanker} &{02m 29s} &{00m 08s}&{03m 38s} &{00m 09s}&{00m 14s}& {00m 01s}\\
DreamRec&01m 31s& 21m 32s&03m 59s& 32m 40s&00m 14s&01m 31s\\
TDM&01m 22s &00m 13s&02m 23s &00m 23s&00m 11s& 00m 01s\\

\toprule
\vspace{-6mm}
\end{tabular}
\end{table}

%% file: 6_conclusion.tex

\section{Conclusion}
In this paper, we propose TDM, a novel approach that simulates extra missing data in diffusion models' guidance signals and extrapolates to address existing missing data in sequential recommendations. By introducing a dual-side Thompson sampling strategy with local and global probability models, TDM can preserve user preference evolution in sequences during simulation. Treating such edited sequences as guidance can achieve diffusion models' consistency regularization. Theoretical analysis and extensive experiments validate the effectiveness of TDM, showcasing its potential to handle missing data issues and improve recommendation performance.
The limitation is that the dynamics of user preferences may exhibit intricate patterns that transcend the capabilities of our dual-side probability models. Future research could benefit from integrating a more sophisticated understanding of user preference evolution into these models. Additionally, the issue of missing data, which can stem from various factors, such as exposure bias or popularity bias, presents an opportunity for targeted simulation to enhance the robustness of diffusion models.

\section{Acknowledgement}
This research is supported by the National Natural Science Foundation of China (92270114, 62302321, 62121002). This research is also supported by the advanced computing resources provided by the Supercomputing Center of the USTC and the Research Center for Intelligent Operations Research at the Hong Kong Polytechnic University.

